\RequirePackage{fix-cm}
\documentclass[nofootinbib,showkeys,superscriptaddress]{revtex4}

\RequirePackage{graphicx}
\usepackage{mathbbol}
\usepackage{graphicx}
\usepackage{amsmath}
\usepackage{dcolumn}
\usepackage{bm}
\begin{document}

\title{Pathway to a Compact SASE FEL Device}


\author{G. Dattoli}
\affiliation{ENEA - Centro Ricerche Frascati, Via Enrico Fermi 45, 00044, Frascati, Rome, Italy}
\thanks{Corresponding Author}
\email{giuseppe.dattoli@enea.it}

\author{E. Di Palma }
\affiliation{ENEA - Centro Ricerche Frascati, Via Enrico Fermi 45, 00044, Frascati, Rome, Italy}

\author{V. Petrillo}
\affiliation{Universit\`{a}  degli Studi di Milano, via Celoria 16, 20133 Milano, Italy}

\author{J.V. Rau}
\affiliation{Istituto di Struttura della Materia, ISM-CNR, Via del Fosso del Cavaliere, 100-00133 Rome, Italy}

\author{E. Sabia}
\affiliation{ENEA - Centro Ricerche Frascati, Via Enrico Fermi 45, 00044, Frascati, Rome, Italy}

\author{I. Spassovsky}
\affiliation{ENEA - Centro Ricerche Frascati, Via Enrico Fermi 45, 00044, Frascati, Rome, Italy}

\author{S.G. Biedron}
\affiliation{Colorado State University}

\author{J. Einstein}
\affiliation{Colorado State University}

\author{S.V. Milton}
\affiliation{Colorado State University}



\begin{abstract}
Newly developed high peak power lasers have opened the possibilities of driving coherent light sources operating with laser plasma accelerated beams and wave undulators. We speculate on the combination of these two concepts and show that the merging of the underlying technologies could lead to new and interesting possibilities to achieve truly compact, coherent radiator devices.
\end{abstract}
\keywords{Free Electron Laser; Self Amplified Spontaneous Emission; Wave Undulator; Radio Frequency Undulator;  Laser Plasma Acceleration; compact FEL.}

\maketitle

\section{Introduction}
\addcontentsline{toc}{chapter}{Introduzione}{}
\markboth{\textsc{Introduzione}}{}
Free Electron Lasers (FELs) operate in many regions of the electromagnetic spectrum and have opened
new and wider perspectives in the applied science; the technologies associated with ultra-small world and
ultrafast phenomena are getting a great deal of benefit within such a context\cite{Web1}. \\
The large size and the non-negligible costs along with the fact that these devices allow few experimental stations,
compared to third generation light source counterparts, the dedicated beam-time is less. The existing and planned facilities
for short wavelengths exploit and foresee the use of large-scale accelerators and long undulators to generate light\cite{Barletta2}.
An effort aimed at reducing both the size and the cost of FEL operating systems could provide a way of making these sources
more accessible in smaller and more numerous laboratories as working tools. \\
For this aim, a number of options have been explored. The efforts have been directed to the reduction of undulator length,
using e.g. electromagnetic devices, hereafter referred as wave undulator (WU), or the size of accelerating system,
by means of the Laser Plasma Acceleration (LPA) mechanism, and more compact undulators\cite{Humieres3}. \\
In terms of the length, the first solution, that of the wave undulator (WU), replaces the standard magnetic undulator's
period that is on the scale of centimeters with that of an "electromagnetic undulator" on the micrometer scale. Using a WU,
this reduces the length of the undulator by a factor of a thousand or more. The LPA option allows accelerating gradients of
tens of GeV/m instead of MeV/m, given by the conventional radio-frequency-based accelerating devices.
This acceleration scheme allows a further reduction on the order of another thousand in terms of length.
We must point out, however, that this solution demands for a laser driver that is not an insignificant device. \\
The merging of the two solutions - a unique coherent photon generating device (undulator) and the LPA acceleration scheme - \cite{Schroeder4} is,
therefore, something worth exploring to drastically reduce the device dimensions and system cost.
This paper contains an analysis of the operation of the specific case of FEL operating in the X-ray (energy and wavelength) region
by using a LPA e-beam and  a WU. With this analysis, we explore the potential of developing a high-gain FEL source scaled to more modest size.
The application of both tools - LPA and WU - could be applied for many configurations of various e-beam energies and/or photon wavelengths.
The variants we will add, however, with respect to previous proposals is the sheared beam geometry \cite{Lawler5} and the use of the same
laser system to provide both the accelerated e-beam and the electromagnetic undulator.

\section{Compact FEL: a few suggestions for the design criteria}

\begin{figure} 
 \begin{minipage}[b]{0.47\textwidth}
 \centering
 \includegraphics[width=.7\textwidth]{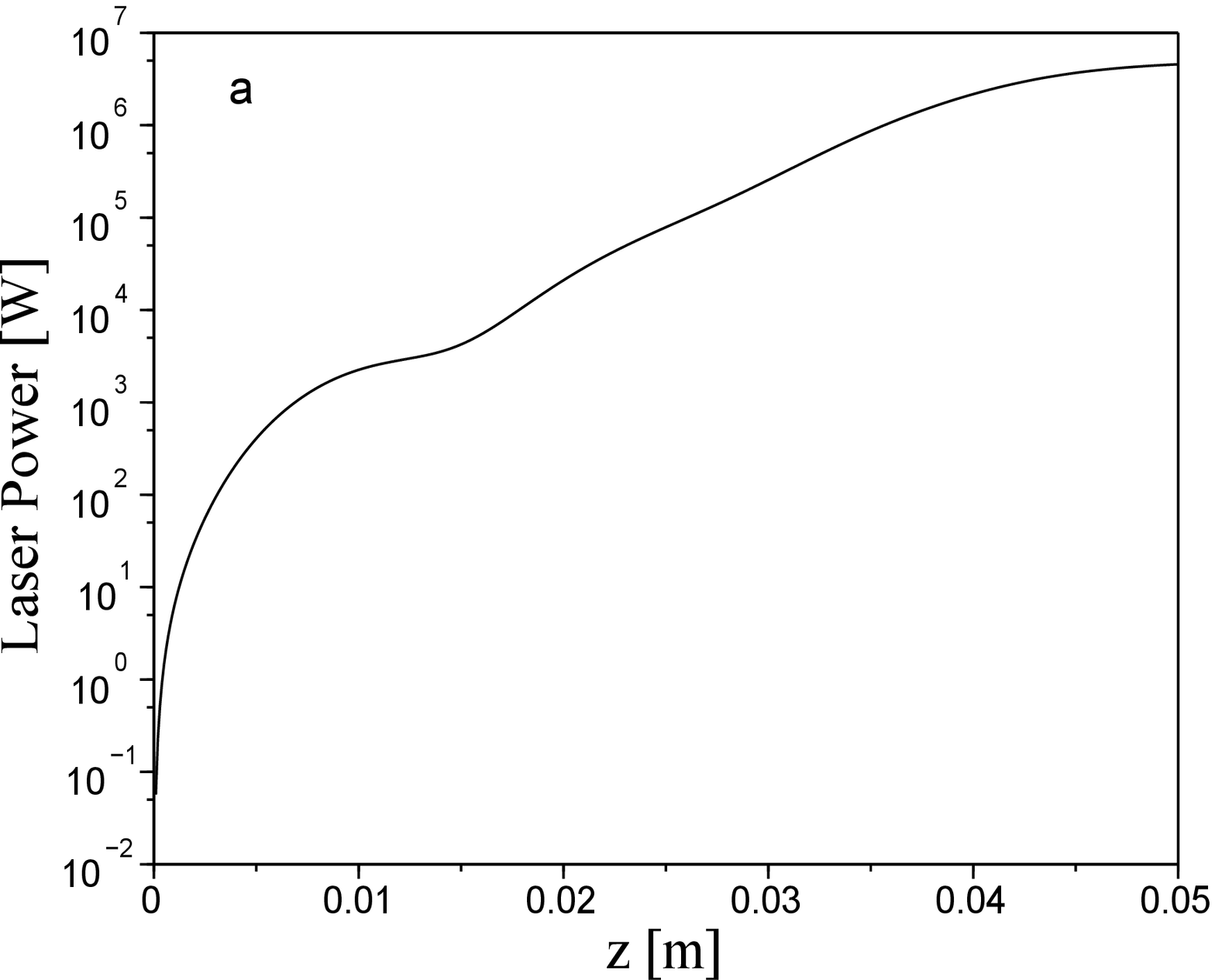}
 \end{minipage}
 \begin{minipage}[b]{0.47\textwidth}
 \centering
 \includegraphics[width=.7\textwidth]{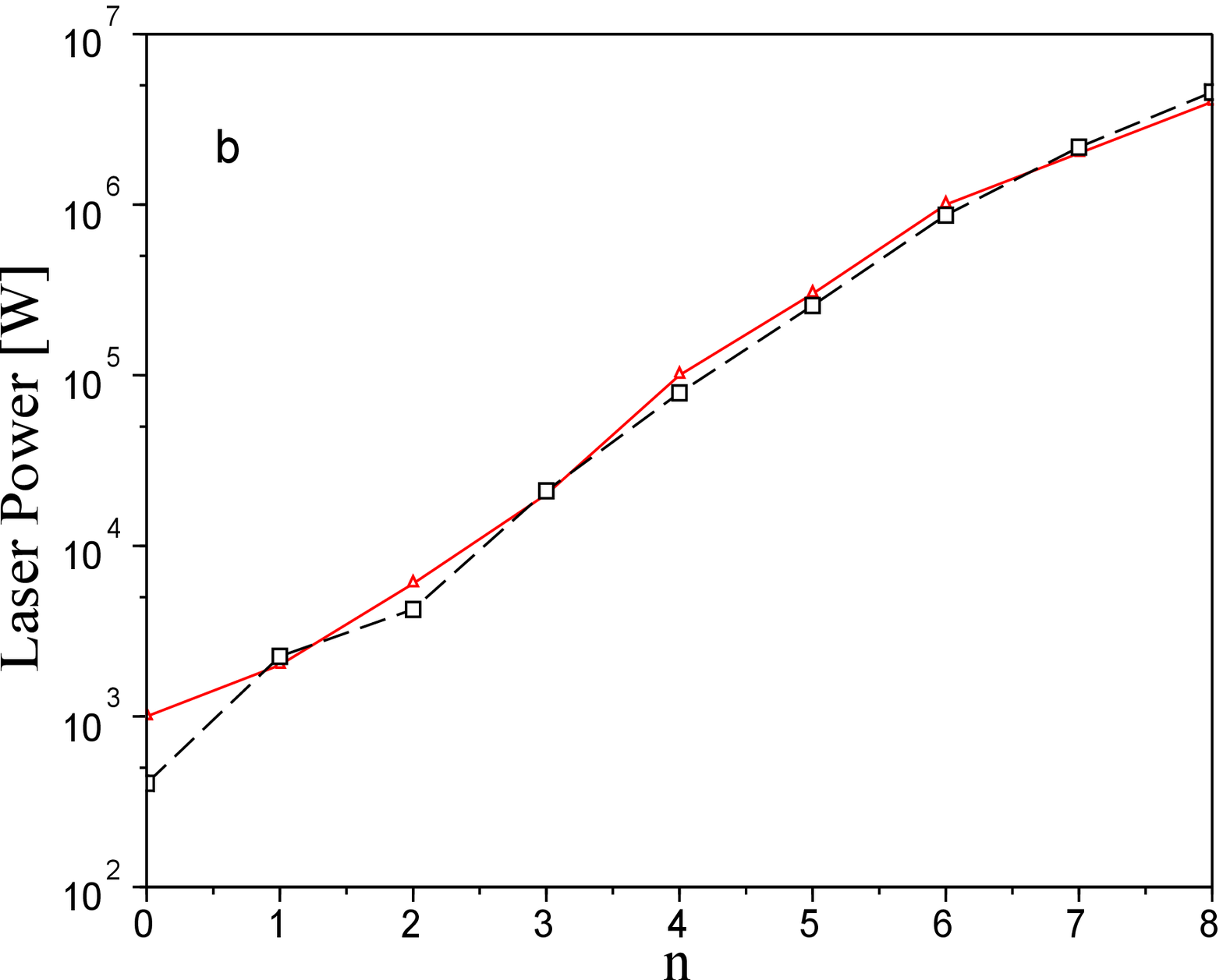}
 \end{minipage}
 \caption{a) Power (W) versus WU longitudinal coordinate $(m)$; b)Comparison between the prediction of GENESIS (triangles)
 and the semi-analytical procedure (boxes), each step corresponds to $5 \cdot 10^{-3}m$.}
 \label{FigPower}
 \end{figure}

We will proceed first by assessing the reliability of the PARSIFEL code \cite{Artioli6}, which will be used in this paper to analyze the
characteristics of the proposed device. This code, based on a combination of semi-analytical formulae and scaling relations,
has been used in the past to analyze the behavior of different FEL facilities operating in various configurations such as oscillators \cite{Dattoli7},
 SASE (Self-Amplified Spontaneous Emission) \cite{Milton8}, and nonlinear harmonic generation \cite{Dattoli9}.
  It is routinely used as an on line bench-mark
 of the SPARC facility \cite{Artioli10}. Very recently an updated version, including the effects of the electron beam transport, has provided
 an important tool for the study of SASE FEL operation with a mismatched electron beam \cite{Artioli10}. \\
The comparison will be made by confronting the PARSIFEL predictions with previous results (either numerical or experimental),
available in the literature. The code, albeit not comparable in terms of completeness with Genesis \cite{Reiche11}, Ginger \cite{Fawley12},
Prometeo \cite{Dattoli13}, Perseo \cite{Giannessi14} or Medusa \cite{Medusa14b} has the advantage of being extremely fast and provides a clear understanding of the
role of various physical parameters entering the laser process. For a full system design, of course, we would perform simulations with several codes
  as well in order to reduce the overall design risk. \\
Once the reliability of the method has been proved, by obtaining already published results concerning the operation in either
the oscillator or SASE, we will discuss the possibility of driving a SASE FEL with an e-beam provided by a $LPA$ process injected into a $WU$. \\
The idea of operating a SASE FEL, using a laser instead of an undulator, emerged since the very beginning of FEL Physics \cite{Pantell15}.
With the advent of powerful laser systems, it is no longer a conceptual proposal \cite{Dattoli16,Bacci17}, and significant efforts have been
devoted towards such realization \cite{Polyanskiy18}. Laser intensities in excess of $10^{26} W/m^2$  have, indeed, been reported \cite{Yanovsky 19},
the upcoming petawatt
lasers aim at intensities of the order of  $10^{27} W/m^2$\cite{Norby20} and intensities of the order of  $10^{28} \div 10^{30} W/m^2$ are
foreseen at the Extreme-Light-Infrastructure
(ELI) \cite{web21} and as described in the US Department of Energy's Laser report \cite{web22}.
Another candidate for operation with a WU is provided by the so called radio-frequency undulators (RFU) \cite{Tantawi23}, which have
the potentials for going to periods around $1 mm$, with reasonably large  $K$ values (comparable with, or even larger than, those
 achievable with a laser). \\
The possibility of designing SASE WU FEL with simple scaling formulae has been recently discussed in \cite{Dattoli24}. We, therefore,
remind that in this type of device, the undulator strength parameter, associated with a $WU$ of intensity $I$ and wavelength $\lambda$,
in practical units reads as:
\begin{equation}
K\approx 0.85 \cdot 10^{-5} \lambda[m] \sqrt{ I\left[\frac{W}{m^2}\right]}
\end{equation}
and that the other quantities (Pierce parameter $\rho$ , gain length $L_g$ and saturated power $P_F$ ), characterizing the
FEL SASE evolution, can be summarized as:
\begin{eqnarray}
\rho&\cong& \frac{8.36 \cdot 10^{-3}}{\gamma} \cdot \left[ J\left[ \frac{A}{m^2}\right] \cdot \left( \lambda[m] K f_b({\xi})  \right)^2 \right]^{1/3} \nonumber \\ \nonumber\\
f_b(\xi)&=&J_0(\xi)-J_1(\xi),\;\;\; \xi=\frac{1}{4}\frac{K^2}{1+\frac{K^2}{2}} \\ \nonumber\\
L_g[m]&\cong& \frac{\lambda[m]}{8 \pi \sqrt{3} \rho} \nonumber\\ \nonumber\\
P_F&\cong& \sqrt{2}\rho P_E \nonumber
\end{eqnarray}
where the parameter $\rho$  has a role of providing the signal growth rate and the transfer efficiency from the electron beam to FEL photons,
 $L_g$  is the gain length of the system and fixes the dimension (length) of the device. $P_F$ is the final laser power at saturation. \\
Finally, the SASE intensity growth vs the "undulator" length, in the longitudinal coordinate $z$ ,
can be specified by the following logistic-like function \cite{Dattoli25}:
\begin{eqnarray}
P(z)&=&\frac{P_0}{9}\frac{B(z)}{1+\frac{P_0}{9 P_f}B(z)} \nonumber \\ \label{eqn:Gain}\\
B(z)&=&2\left[cosh\left( \frac{z}{L_g} \right)- e^{-\frac{z}{2L_g} }cos\left(\frac{\pi}{3}+\frac{\sqrt{3}z}{L_g} \right) -  e^{\frac{z}{2L_g} }cos\left(\frac{\pi}{3}-\frac{\sqrt{3}z}{L_g} \right)    \right] \nonumber
\end{eqnarray}
In Fig.\ref{FigPower}a, we report the power growth of a WU FEL operating at $\lambda\cong 1.35 nm$  and using a $CO_2$ laser, as undulator driver,
with a power corresponding to $K\cong0.3$,  $\left(I\left[ \frac{W}{m^2}\right] \cong 9 \cdot 10^{18} \right)$. In Fig. \ref{FigPower}b,
we compare the results obtained by the GENESIS and those from the eq.\ref{eqn:Gain}, respectively. The comparison is good, at least for
the present purposes. Any discrepancies are related to the differences in the $1D$ and $3D$ representations of the interaction and
also due to the fact that the GENESIS starts from the random noise in the electron beam. \\
The previous analysis does not include, however, the longitudinal distribution of the optical pulse, which may be responsible
for the corresponding dependence of the K parameter on the pulse profile.
The problem should be carefully taken into account in the simulations. Just to give an idea of the importance of the problems
involved in, we use a fairly simple argument useful to fix the operating parameters.

If we assume that the optical pulse has a Gaussian shape with rms value sz we impose the following condition
\begin{equation}
\frac{K_0-K_i}{K_0}=1-e^{\frac{-\tau_i^2}{4 \sigma_{\tau}^2}}\approx \rho
\label{eqn-new}
\end{equation}
which states that over the pulse length, from the beginning of the FEL interaction to the maximum of the pulse,
the inhomogeneous broadening effects, induced by the longitudinal intensity shapes,  are within the gain bandwidth.
From eq. \ref{eqn-new} we find that
\begin{equation}
\tau_i\approx 2 \sqrt{\rho}\sigma_{\tau}
\label{eqn-new1}
\end{equation}
Imposing that $\tau_i$ corresponds to $1/2$ of the saturation length, we obtain the condition
\begin{equation}
\sigma_{\tau\approx} 4 \sqrt{\rho} \frac{L_s}{c}
\label{eqn-new2}
\end{equation}
If we accept a saturation length of $0.005m$ with a $\rho=5\cdot10^{-4}$ we should demand for a r.m.s  pulse duration $\sigma_{\tau}\approx15ps$.
If we require a $K$ value around $0.3$, the necessary intensity per pulse should bring a very high energy density (in excess of $10^8 J/m^2$ ),
which will eventually be realized in the next generation of high power lasers.

\begin{figure}[h]
 \centering
\includegraphics[width=.5\textwidth]{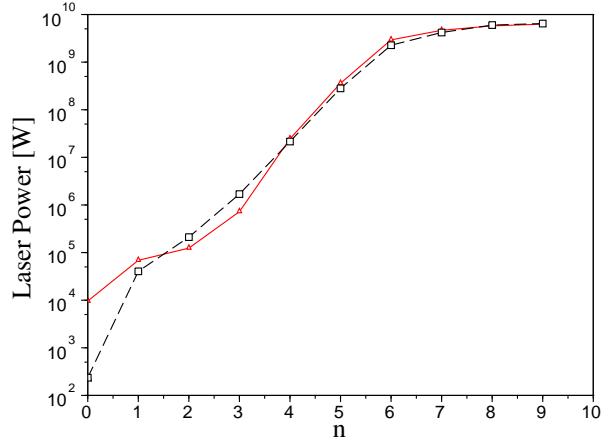}
 \caption{Power (W) versus longitudinal coordinate: comparison between the predictions of the Genesis (Triangles)
 and of the Analytical model (Boxes). Each step $n$ corresponds approximately to $0.5 m$.}
 \label{fig2}
\end{figure}

LPA driven accelerators have made significant progress during the last years. Beams with energies comparable to those of
conventional synchrotrons have been reached. Albeit the beam qualities are still insufficient, in terms of energy dispersion,
furthermore shot-to-shot stability control is presently rather poor, it is however hoped that, in the near term, stable laser
accelerated beams will be developped for applications. \\
One of these, LPA based SASE FEL, has been proposed \cite{Maier26}, and its principle has been tested \cite{Seggenbrock27}.
The preliminary results show that both radiation and tunability can be achieved from the beam emission inside the undulator. \\
The set of parameters reported in Tab. \ref{tab1} is not too far from those achievable in the near future. They have, therefore,
been exploited in start-to-end simulations modeling LPA SASE FEL with the Genesis \cite{Hooker28}. In Fig.\ref{fig2}, we have also included
the PARSIFEL predictions, and the comparison yields more than a reasonable agreement. We can, therefore, consider the use
of our semi-analytical approach, reliable for the purposes of a quick evaluation of the FEL performances within different configurations. \\
\begin{table}[h]
\centering
\begin{tabular}{l}
  $E\equiv$ electron beam energy$=400MeV$\\
  $Q\equiv$ charge per bunch$=250pC$ \\
  $\tau\equiv$r.m.s. bunch length$=10fs$ \\
  $\varepsilon\equiv$ Normalized Emittance $=1mm\cdot mrad$ \\
  $\lambda_u\equiv$ undulator period$=1.5 cm$ \\
  $K\equiv$undulator strength parameter$=2$ \\
\end{tabular}
\caption{Set of parameters used to simulate the LPA SASE FEL with Genesis.}
\label{tab1}
\end{table}

\begin{figure}[h]
 \centering
 \includegraphics[width=.5\textwidth]{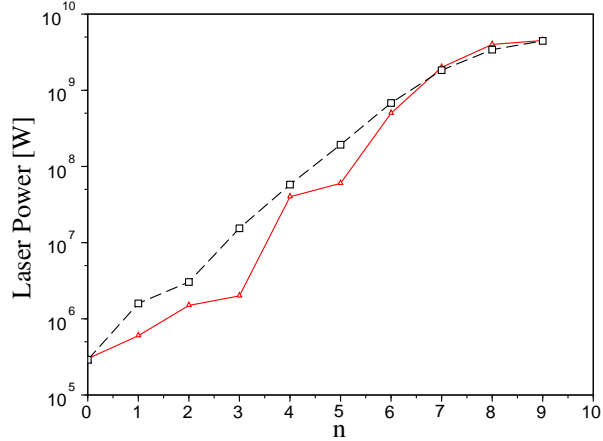}
 \caption{Same as Fig. \ref{fig2}, but for the operation at $10nm$ (each step is $5 m$) ($E=750MeV, \; K=1.9,\;\lambda_u=1.5cm$ ).}
 \label{fig3}
\end{figure}

To avoid misunderstandings and for the sake of completeness we add a further comparison (Fig. \ref{fig3}),
relevant to the operation at $10 nm$ with the same beam parameters of Fig. \ref{fig2}, but with a slightly different undulator.
The comparison is not that striking as in the previous cases, but still acceptable. \\

\begin{figure}[h]
 \centering
 \includegraphics[width=.5\textwidth]{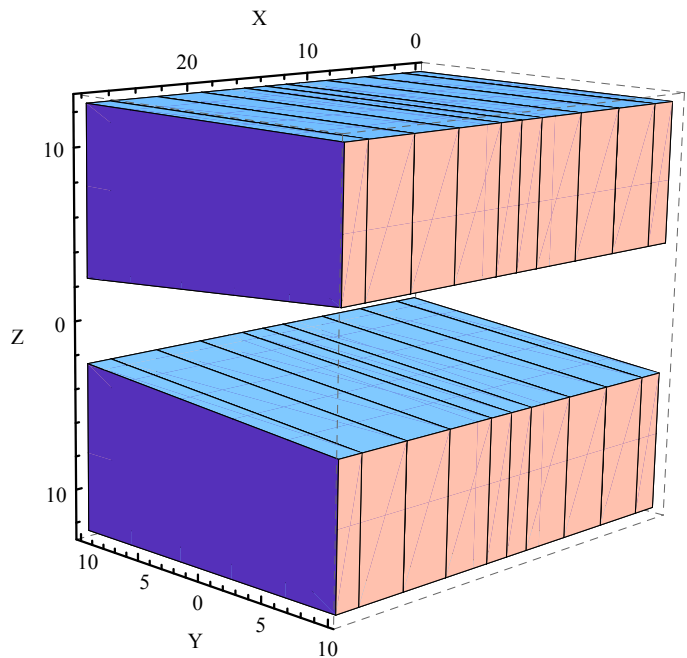}
 \caption{Scheme of a transverse gradient undulator.}
 \label{fig4}
\end{figure}

In the case of SASE FEL driven by LPA beams the electron beam is affected by a large e-beam energy spread, which, as it is well known, induces an
inhomogeneous broadening causing a decrease in the gain and a consequent increase in the saturation length. Such a
spoiling effect could be counteracted by a large value of the Pierce parameter that for the specific case reported in Table \ref{tab1},
is $\rho_o\cong 5.85 \cdot 10^{-3}$ (the subscript $o$ stands for homogeneous) the inclusion of the inhomogeneous
broadening correction \cite{Dattoli25}
\begin{equation}
\rho=\frac{\rho_0}{\left(1+0.185 \cdot \frac{\sqrt{3}}{2} \widetilde{\mu}_{\varepsilon}^2 \right)},\;  \widetilde{\mu}_{\varepsilon}=\frac{2 \sigma_{\varepsilon}}{\rho_o}
\end{equation}
determines a reduction of more than $50\%\;(\rho\cong2 \cdot 10^{-3} )$. For larger energy spread and lower beam current, different
solutions should be devised to mitigate  such a detrimental contribution. One possibility is the use of the so called transverse
gradient undulators (TGU), which can both counteract the effect due to energy distribution and provide a more efficient operation.
In TGU devices a transverse tapering of the magnetic
 field is induced, providing an energy dispersion in the tapering direction (see Fig.\ref{fig4}). \\

\begin{figure}[h]
 \centering
 \includegraphics[width=.5\textwidth]{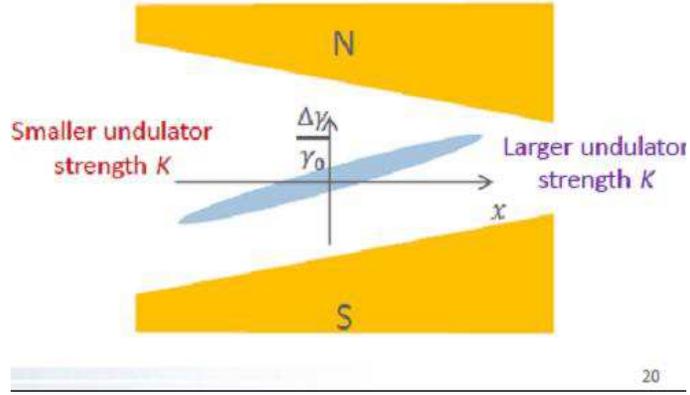}
 \caption{Scheme of a transverse gradient undulator.}
 \label{fig4tris}
\end{figure}

To better clarify the working principle of the device we make reference to Fig.\ref{fig4tris}  where we have reported the transverse profile
of an undulator along with the associated tapering, realized by canting its upper and lower faces.\\
The LPA produced e-beam are characterized by an energy transverse position correlation, therefore, the tapering should be realized
in such a way that the portion of the beam with lower energy intercepts a region of the undulator with weaker field intensity and
viceversa, so that all the bunch region radiate the same wave length, thus cancelling the inhomogeneous broadening effects.\\
If the energy and the strength parameters are specified through the relations
\begin{eqnarray}
\gamma(x)&=&\gamma_0(1+D^{-1}x) \nonumber \\\label{eqnA} \\
K(x)&=&(1+ \kappa x) \nonumber
\end{eqnarray}
with $D$ and $\kappa$ being the energy dispersion function and undulator gradient, respectively, we end up with the
following condition necessary to cancel the inhomogeneous broadening effects
\begin{equation}
D \kappa=\frac{2+K_0^2}{K_0^2}
\end{equation}

\begin{figure}[h]
 \centering
 \includegraphics[width=.5\textwidth]{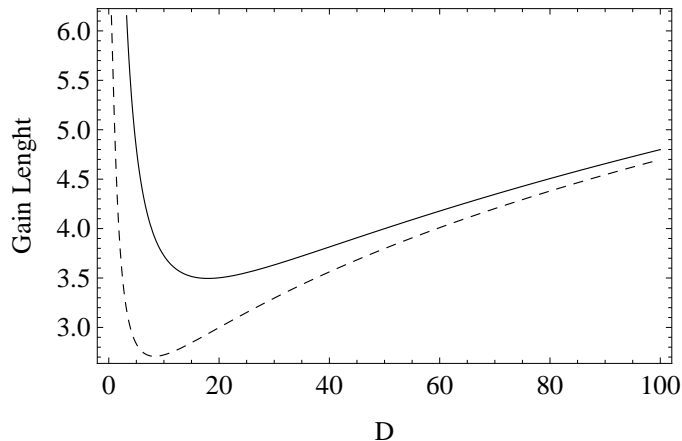}
 \caption{Gain length vs. $D$ for $\widetilde{\mu}_{\varepsilon}=6$(dot) and  $\widetilde{\mu}_{\varepsilon}=10$ (continuous).}
 \label{fig2bis}
\end{figure}

The presence of a non- zero dispersion function provides an increase of the transverse dimensions according to
\begin{equation}
\sigma_T=\sqrt{\sigma_x^2+(D \sigma_{\epsilon})^2}
\end{equation}
Which is associated with a reduction of the energy spread (due to the conservation of transverse longitudinal phase space) according to
\begin{eqnarray}
\sigma_{\epsilon}'=\frac{\sigma_x}{\sigma_T}\sigma_{\epsilon}
\label{eqnD}
\end{eqnarray}
As we already remarked the energy spread induces an increase of the gain length according to the equation
\begin{equation}
L_g=L_g^0\left[ 1+0.185 \frac{\sqrt{3}}{2} \widetilde{\mu}_{\varepsilon}^2 \right]
\end{equation}
Therefore, by redefining the energy spread through eq. \ref{eqnD} we should be able to quantify the beneficial effect of the transverse tapering.
The beneficial effect of the reduction of the relative energy spread is counteracted by the increase of beam section, determining
a decreasing of the current density, which in turn determines a reduction of the Pierce parameter as reported below
\begin{equation}
\overline{\rho}=\rho_0 \left(\frac{\sigma_x}{\sigma_T} \right)^{1/3}
\end{equation}
By including either effects in the definition of the gain length, we end up with
\begin{equation}
L_g=L_g^0\left[ \left(\frac{\sigma_x}{\sigma_T} \right)^{-1/3}+0.64 \left(\frac{\overline{\sigma}_{\epsilon} }{\overline{\rho} } \right)^2 \left( \frac{\sigma_x}{\sigma_T} \right)^{-1} \right]
\end{equation}
The gain length can accordingly be written as
\begin{eqnarray}
L_g(\Delta,\widetilde{\mu}_{\varepsilon})&=&L_g^0 \left[ \left( 1+\Delta^2 \right)^{1/6}+0.16 \widetilde{\mu}_{\varepsilon}^2 \left( 1+\Delta^2 \right)^{-1/2} \right] \nonumber\\ \\
\Delta&=&\frac{D \sigma_{\epsilon} }{\sigma_x} \nonumber
\end{eqnarray}
In Fig. \ref{fig2bis} we have reported $L_g/L_g^0$ vs. $\Delta$, for different values of $\widetilde{\mu}_{\varepsilon}$.\\
The function exhibits a minimum, depending on the values of the "natural" inhomogeneous broadening,  in correspondence of such a minimum,
denoted by $\Delta^*$  we can fix the optimum of the dispersion function as
\begin{equation}
D^*\approx  \frac{\sigma_x}{\sigma_{\epsilon}} \Delta^*
\end{equation}

\begin{figure}[h]
 \centering
 \includegraphics[width=.5\textwidth]{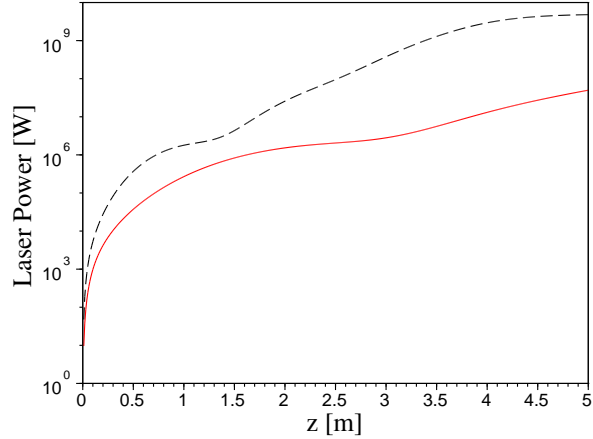}
 \caption{FEL operation with and without TGU, the same parameters of Fig. \ref{fig2};
 the transverse tapering parameters have been chosen to halve the effect of the inhomogeneous broadening
 (for further comments see the third of refs. \cite{Huang29}).}
 \label{fig5}
\end{figure}

The minimum of the gain length occurs in correspondence of
\begin{equation}
\Delta^* \approx \sqrt{0.333\widetilde{\mu}_{\varepsilon}^3-1 }
\end{equation}
which yields, for the dispersion function
\begin{equation}
D^*\approx \frac{\sigma_x}{\sigma_{\epsilon}} \sqrt{0.333\widetilde{\mu}_{\varepsilon}^3-1 }
\label{eqnM}
\end{equation}
for $\mu_{\epsilon} \gg 1$  the eq. \ref{eqnM} reduce to
\begin{equation}
D^*\approx 1.15 \frac{\sigma_x}{\rho_0}\sqrt{\widetilde{\mu}_{\varepsilon}}
\end{equation}
The optimum gain length writes
\begin{equation}
L_g(\Delta^*,\widetilde{\mu}_{\varepsilon})\approx 1.1 \sqrt{\widetilde{\mu}_{\varepsilon}} L_g^0
\end{equation}
and can be exploited to get a first idea of how the various terms should be combined in the design of a FEL device operating
with a large energy electron beam to mitigate the relevant negative effects.\\
To illustrate the possible perverse effect of the energy spread on the gain length, we pick out, as reference parameters, those considered in
ref. \cite{Huang28bis} and reported in Tab.\ref{tab1bis}. \\
\begin{table}[h]
\centering
\begin{tabular}{ll}
  Undulator parameter   & $K=2$\\
  Undulator period      & $\lambda_u=1cm$ \\
  Beam energy           & $E=1GeV$ \\
  Resonant wavelength   & $\lambda\approx 1nm$ \\
  Peak current          &   $I\approx 10kA$ \\
  Energy spread         &  $\sigma_{\epsilon}\approx 10^{-2}$ \\
  Normalized emittances &  $\gamma \epsilon_{x,y}\approx 1 \mu mrad$     \\
  Horizontal and vertical size  & $\sigma_{x,y}\approx 11.3 \mu m$\\
  FEL parameter         &   $\rho \approx 6 \cdot 10^{-3}$ \\
\end{tabular}
\caption{Plasma beam and FEL parameters.}
\label{tab1bis}
\end{table} \\
With the above list we find $\widetilde{\mu}_{\varepsilon}\approx 3.3$, which determines an increase of the gain length with respect to
the homogeneous case of about a factor 3.\\
The use of the previous optimization criteria applied to the case of Tab.\ref{tab1bis},  determines an increase of the gain length of a factor 2 only,
with an associated dispersion function $D^*\approx 4\cdot 10^{-3} m$.\\
The analysis developed so far is not dissimilar from that suggested in ref.\cite{Huang28bis}, apart from unessential details due to the different
parametrization formulae we used here (for further comments see ref. \cite{Ciocci28tris}).
The line shift due to energy and undulator tapering are the same, but opposite in sign and cancel, therefore,
each other, thus compensating the effect of the energy spread \cite{Huang29}. \\
A comparison between TGU and conventional undulator operation is shown in Fig. \ref{fig5}, where we have considered the
same parameters of Fig. \ref{fig2}.

\begin{figure}[h]
 \centering
 \includegraphics[width=.5\textwidth]{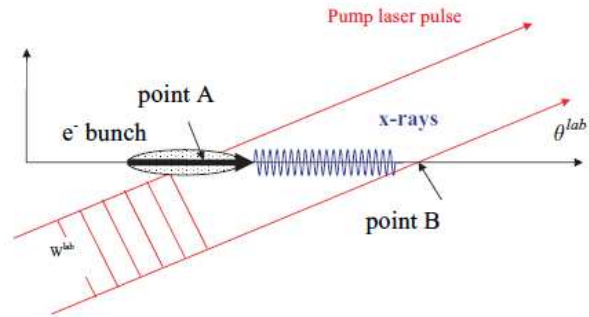}
 \caption{Sheared Configuration from ref.\cite{Fowler31}}
 \label{fig6}
\end{figure}

\begin{figure}[h]
 \centering
 \includegraphics[width=.5\textwidth]{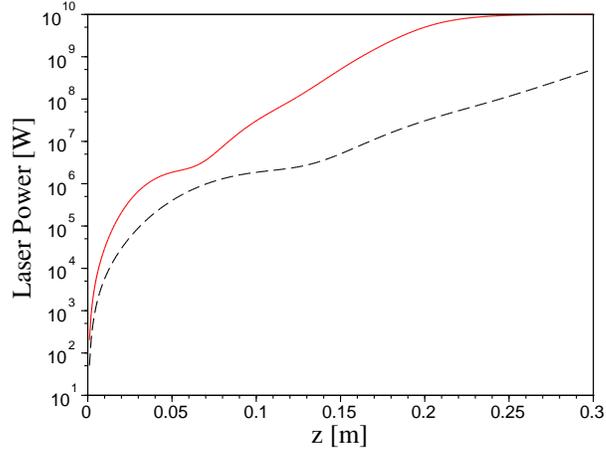}
 \caption{Power versus longitudinal coordinate $(m)$ for a LPA WU FEL.}
 \label{fig7}
\end{figure}

Having sufficient reasons to be confident in our computational tool, we can discuss a scheme of a FEL
wave undulator driven by a plasma accelerated e-beam. \\
In the following, we will assume that the accelerating mechanism relies on the so-called bubble accelerating regime.
We use the scaling formulae derived by Pukhov and Gordienko \cite{Pukhov30} to fix the parameters of the LPA accelerating system.
Accordingly, we will write the beam energy $E_e$  and the number of accelerated electrons $N_e$  in terms of the laser parameters as:
\begin{eqnarray}
E_e&\cong& \frac{0.65}{2\pi} m_ec^2  \sqrt{\frac{P_L}{P^{\ast}}} k \sigma_L  \nonumber\\ \label{eqn5} \\
N_e&\cong& \delta \frac{1.8}{k r_e}  \sqrt{\frac{P_L}{P^{\ast}}}   \nonumber
\end{eqnarray}
where $P_L$  is the laser power, $P^{\ast}\cong8.5GW$  is the threshold power for relativistic acceleration, $k=2\pi/\lambda$ is the laser
wave vector, $\sigma_L$  the length of the pulse and  $r_e$ the electron classical radius. The parameter $\delta$  depends on the
energy distribution and on the energy spread, one can reasonably accept for a safe  FEL operation. In the following,
it will be treated as a free parameter. \\
The acceleration length is linked to the laser parameters by the following equations:
\begin{eqnarray}
L_{acc}&\cong& 0.7 \frac{\sigma_L}{\lambda}Z_R  \nonumber \\ \label{eqn6} \\
Z_R&=& \frac{\pi R^2}{\lambda}  \nonumber
\end{eqnarray}
$Z_r$ being the Rayleigh length and $R$  the bubble radius, furthermore, the condition for the PLA requires $\sigma_L \leq\leq R$.
In the following, we will assume that the electron bunch has the same length of the laser pulse, therefore, from the second of eqs. \ref{eqn5},
we find that the associated beam current is:
\begin{eqnarray}
\widehat{I}&\cong& \delta \frac{1.8}{k \sigma_L} \sqrt{\frac{P_L}{P^{\ast}}} I_0  \nonumber \\ \label{eqn7} \\
I_0&=& \frac{e c}{r_e}\nonumber
\end{eqnarray}
where $I_0=1.7 \cdot 10^{4}A$ is the Alfv\`{e}n current. The beam power can, therefore, be written as $(P^{\ast}=\frac{m_e c^2}{e} I_0 )$:
\begin{equation}
P_E\cong\delta \cdot P_L
\label{eqn8}
\end{equation}
We can now use the previous formulae to get an idea of the numbers involved in the operation of a LPA WU SASE FEL.
By keeping $\sigma_L=\chi R$, $\chi\leq 1$  we can link the radius of the accelerating bubble to the accelerating length according to the identity:
\begin{equation}
R\cong \sqrt[3]{\frac{L_{acc}\lambda^2}{0.7 \chi \pi}}
\end{equation}
Thus, finding for the laser power necessary to bring the system to the energy $E_e$
\begin{eqnarray}
P_L&\cong& 2 \chi^{-4/3} \left( \frac{\pi \lambda}{L_{acc}} \right)^{2/3} \gamma_e^2 P^{\ast}  \nonumber\\ \label{eqn10} \\
\gamma_e&=&  \frac{E_e}{m_e c^2}\nonumber
\end{eqnarray}
Finally, the achievable peak current is:
\begin{equation}
\widehat{I}\cong 0.8\delta \sqrt[3]{\frac{\lambda}{\chi^2 \pi^2 L_{acc}}} \sqrt{\frac{P_L}{P^{\ast}}} I_0
\end{equation}
We can now use the previous relations to derive the laser characteristics. We fix the following acceleration distance
 $L_{acc}\cong 2 \cdot 10^{4} \lambda, \; (\lambda \cong 1 \mu m)$, to reach a maximum beam energy of $200MeV$, which, assuming  $\chi\cong0.5$,
 $\delta=1$, yields the following set of parameters:
\[
R\cong2.6 \cdot 10^{-5}m, \; P\cong1.9\cdot10^{13} W, \; \tau_L\cong40 fs,\; \widehat{I}\cong1.03\cdot I_0
\]
The current is quite large, but such a value can be exploited to compensate the effect of the e-beam energy spread, which might be quite large.\\
Let us now consider a WU provided by a laser with the characteristics used to derive the plots in Fig. \ref{FigPower}.
It is evident that if we use a head-on collision between such a laser and the PL produced beam, the scattered
process determines photons in the range of gamma rays. \\
If we use instead the sheared laser configuration suggested in ref. \cite{Fowler31} and reported in Fig. \ref{fig6}
in this scattering geometry the electron and laser beams are nearly co-propagating, the induced Doppler shift is,
therefore, reduced to yield soft, rather than hard $X-rays$ or even $\gamma-rays$. The scattered wavelength is, indeed:
\begin{eqnarray}
\lambda_X&=& \frac{\lambda(1-\beta)}{1-\beta cos(\theta_{lab})}\cong \frac{\lambda_F}{sin\left( \frac{1}{2} \theta_{lab}\right)^2}   \nonumber \\ \label{eqn12}
\lambda_F&=& \frac{\lambda}{4 \gamma^2}\nonumber
\end{eqnarray}
If we use the parameters reported in Table \ref{tab1} ($\lambda\cong750 nm$  and a shear angle with $cos(\theta_{lab})=0.9983$),
we obtain $\lambda_X\cong1.4nm$. The length of the interaction region can be fixed by the use of quite a simple argument.
To have an estimation of the orders of magnitudes involved we note that the "equivalent undulator" period is:
\begin{equation}
\lambda_u^{SH}\cong \frac{\lambda}{2 sin\left( \frac{\theta_{lab}}{2}\right)}
\end{equation}
(about $13 \mu m$ for the parameters reported in Table \ref{tab2}) the corresponding gain length will, therefore, be expressed by
\[
L_g=\frac{\lambda_u^{SH}}{4 \pi \sqrt{3}\rho}
\]
by assuming that the saturation length is $L_S\cong 20L_g$ and that $\rho\cong 10^{-3}$, we obtain an interaction length of less than $0.012 m$.\\
Regarding the Pierce parameter $\rho$ we will write it in the form (valid for a circularly polarized mode)
\[
\rho_0=\left[ \frac{K^2 \gamma \lambda_X^2 r_e n_{lab}}{4 \pi}  \right]^{1/3}
\]
where $n_{lab}$ is the e-beam density in the lab-frame, which is linked to the current by
\[
n_{lab}=\frac{I}{2 \pi c \sigma_x \sigma_y}
\]
If we consider a round beam with $50\mu m$ transverse length, we find  for $n_{lab}$ the value reported in Tab. \ref{tab2}, which,
along with the other values therein reported, yields $\rho_0\cong7.14 \cdot 10^{-3}$, and the inclusion of the effects of the
inhomogeneous broadening yields a reduction of $50\%$. With these numbers we obtain the power growth reported in Fig. \ref{fig7},
in which the dot curve is relevant to the inclusion of other effects determining a further increase (another factor of $2$) of the gain length. \\
\begin{table}[h]
\centering
\begin{tabular}{l}
  $\lambda_L\equiv$ laser wave length $\cong750 nm$\\
  $I_L\equiv$ laser intensity $\cong 10^{22} \frac{W}{m^2}$ \\
  $n_{lab}\equiv$ lab-frame e-b density $\cong 4.3 \cdot 10^{24} m^{-3}$ \\
  $K=0.435$ \\
  $\varepsilon\equiv$ normalized emittance $=1mm\cdot mrad$ \\
  $\lambda_u\equiv$ equivalent undulator period $\cong13 \mu m$ \\
\end{tabular}
\caption{}
\label{tab2}
\end{table}

An interesting aspect of this scheme is the possibility of exploiting one laser beam only,
split into two parts, one being exploited to generate the PLA beam and the other acting as sheared WU. \\
It is worth to note that, since $K^2\propto I$ then we have $\rho \propto \sqrt[3]{I}$, the SASE FEL power scales, therefore,
as $P_F\propto \sqrt[3]{I} \cdot P_L \propto I^{4/3}$. \\
The bare essence of the process we have described is, therefore, that of a parametric up conversion, in which the laser
radiation is converted into that of shorter wavelength, by means of a non-linear device combining the PLA and sheared laser
beam undulator mechanisms. \\
Before closing this paper we want to mention the possibility of exploiting a RFU undulator, namely a different flavor of WU.
The present status allows the achievement of short periods ($1.4 cm$) and large equivalent on axis magnetic field
(on the order of $6 kG$)\cite{Tantawi23}. They are powered by a microwave, provided by an X-band klystron, the field strength is associated
with the injected power by the relation \cite{Shumail32}:
\begin{equation}
K^2 \cong \frac{\sqrt{\lambda_u[cm]}}{2.46} N_u^{-2/3} P[MW]
\end{equation}
The achievable values in terms of $K$ values and of period length cannot, at moment, suggest application for fourth generation
synchrotron radiation devices. They are, however, interesting for low gain oscillators and the experimental results of ref.\cite{web22} support
this possibility. One of the most appealing conclusions of \cite{web22} is the use of a configuration  in which a beam of electrons
undergoing a bunching in a first undulator is injected into a second undulator, where it undergoes a "seeded" coherent harmonic generation (SCHG).
 Even though the adjective "seeded" in not fully appropriated, since the emission is due to a harmonic bunching induced by the
 seeding in the first part of the undulator. The scheme is reminiscent of some theoretical proposals put forward in the
 past \cite{Dattoli33,stupakov34}. The most significant breakthrough being, however, that the second undulator, where the emission actually
 occurs, is a RFU. \\
The results reported in the experiment suggest that this undulator can be exploited to drive an oscillator
(we will provide a detailed analysis elsewhere). We only note here that the achievable values of undulator strength and period may
allow the operation in the UV region with an e-beam of modest energy. Such an oscillator could have interesting properties, associated
with the possibility of piloting the RFU polarizations. The technical issues as the stabilization of $K$ within a band of
$\frac{\Delta K}{K}\ll \frac{1}{4 N_u}$ will be discussed in a separate publication. \\
A further possibility is offered by these type of devices, if exploited together with other undulator devices.
A promising scheme is the use of RFU undulator with low K value coupled with the second conventional undulator adjusted on an harmonics of the RFU.
If a powerful seed laser, resonant with the RF undulator  is exploited, the large seed power is capable of inducing bunching at higher harmonics,
 in spite of the low gain associated with the small $K$ ($<0.1$) value of the RFU. The harmonics supported by the conventional undulator
 can then grow due to the FEL mechanism. We have checked the feasibility of such a scheme, which is even more appealing because
 the RF undulator is not particularly demanding in terms of K-values and thus of power to be handled. A power of $6 MW$ is, indeed,
 needed for RFU with $K=0.1$, a wavelength of $10 cm$ and $N=50$.

\section{Conclusions}
In this paper, we have used simple analytical means based on (reliable) scaling formulae to analyze a number of possible solutions providing a kind of road map for the realization of compact short
wavelength FEL source. From our analysis it emerges that pivoting elements of the discussion are plasma accelerated beams
and laser wave undulators. \\
On the other side, the idea of exploiting TGU undulator schemes is extremely interesting and should be pursued, even though a
careful design modeling the effect of the sextupolar terms on the laser dynamics is needed. These terms may be
responsible for an additional inhomogeneous broadening associated with the transverse electron distribution, which may nullify all
the benefits induced by the TGU design. \\
We believe that the use of RF undulators could provide an extremely helpful tool, also if exploited at the present state of the art.
They may offer, indeed, a wealth of possibilities, including the superposition with an ordinary magnetic undulator,
combined in such a way that a bi-harmonic geometry \cite{Dattoli35}  is realized or to guarantee mild transverse and longitudinal tapering,
as it will be shown elsewhere.


\begin{thebibliography}{}
\bibitem{Web1}
See for example lightsources.org
\bibitem{Barletta2}
W. A. Barletta, et al, \emph{Free Electron Lasers: Present status and future challenges},  NIM A618, p. 69 (2010).
\bibitem{Humieres3}
E. d'Humieres and P. Balcou, \emph{Compact X-FEL Schemes, Free Electron Lasers}, Ed. by Sandor Varro, ISBN 978-953-51-0279-3, Published: March 14, 2012 under CC BY.
\bibitem{Schroeder4}
G. B. Schroeder et al.,\emph{Proceedings of the FEL 2012},Conference Nara, Japan (2012).
\bibitem{Lawler5}
 J E Lawler et al,J. Phys. D: Appl. Phys. 46 325501 (2013) doi:10.1088/0022-3727/46/32/325501.
\bibitem{Artioli6}
 M. Artioli, G. Dattoli, P. L. Ottaviani, S. Pagnutti,Rivista ENEA, Energia, Ambiente e Innovazione, 3/2012
\bibitem{Dattoli7}
G. Dattoli et al.,Nucl. Instr. Meth. A, Volume 545, Issue 1-2, p. 475-479 (2005)
\bibitem{Milton8}
S. V.  Milton et al., Phys. Rev. Lett. 85,998 (2000).\\
P. Emma et al., Nature Photonics 4, 641 (2010).
\bibitem{Dattoli9}
G. Dattoli, P. L. Ottaviani and S. Pagnutti, J. Appl. Phys. 97, 113102 (2005).\\
S. G. Biedron et al., Phys. Rev. STAB 5, 030701 (2002).
\bibitem{Artioli10}
M. Artioli, G. Dattoli, P. L. Ottaviani and S. Pagnutti,  \emph{Virtual Laboratory and Computer Aided Design for Free Electron Lasers outline and simulation}, Energia Ambiente E Innovazione, n. 3 May-June (2012).\\
L. Lazzarino et al.PRSTAB to be published
\bibitem{Reiche11}
S. Reiche, Nuclear Instruments and Methods in Physics Research A 429 (1999) 243
\bibitem{Fawley12}
W. M. Fawley, PRSTAB 5, 070701 (2002)
\bibitem{Dattoli13}
G. Dattoli, M. Galli, and P.L. Ottaviani, \emph{1-Dimensional Simulation of FEL Including  High Gain Regime Saturation, Prebunching, and Harmonic Generation}, ENEA internal report RT/INN/93/09 (1993).
\bibitem{Giannessi14}
L. Giannessi, Proc. 2006 FEL Conf. (JACOW) 91-94 (2006)
\bibitem{Medusa14b}
H.P. Freund, P.G. O'Shea, S.G. Bedron, Nucl. Instrum Meth. A 528 (2004) 44-47
\bibitem{Pantell15}
R.H. Pantell, G. Soncini, H.E. Puthoff, IEEE Journal of Quantum Electronics 4 (1968) 905; \\
P. Dobsiach, P. Meystre and M. O. Scully, IEEE J.QE 19 (1983) 1812; \\
F. Ciocci, G. Dattoli and J. Walsh, Nucl. Instrum. Meth. A 237 (1985) 401; \\
J. Gea Banacloche, G. T. Moore, R. R. Schlicher, M. O. Scully and H. Walther, IEEE J. QE 23 (1987) 1558 ;\\
Cha Mei Tang, B. Hafizi and S. K. Ride, Nucl. Instrum Meth. A 331 (1993) 371.
\bibitem{Dattoli16}
G. Dattoli, T. Letardi and L.R. Vazquez, Phys. Lett. A 237 (1999) 26.
\bibitem{Bacci17}
A. Bacci, M. Ferrario, C. Maroli, V. Petrillo and L. Serafini, Phys. Rev. Special Topics (Accelerators and Beams) 9 (2006) 060704; \\
C. Maroli et al., Proceedings of the FEL 2007 Novosibirsk Russia.
\bibitem{Polyanskiy18}
M. N. Polyanskiy, I. V. Pogorelsky and V. Yakimenko Optics Express 7717, Vol. 19, April 11 (2011);
V. Yakimenko, \emph{CO2 Laser Based undulator for a compact SASE FEL},  proceedings of  Laser and Plasma Accelerators Workshop 2011 Brookhaven (New Jork) June 20-21 (2011)
\bibitem{Yanovsky 19}
V. Yanovsky et al., Opt. Express 16, 2109 (2008).
\bibitem{Norby20}
J. Norby, Laser Focus World, January 1 (2005).
\bibitem{web21}
http://www.extreme-light-infrastructure.eu/.
\bibitem{web22}
http://www.acceleratorsamerica.org/report/index.html; \\ http://science.energy.gov/~/media/hep/pdf/accelerator-rd-stewardship/Lasers\_for\_Accelerators\_Report\_Final.pdf
\bibitem{Tantawi23}
S. Tantawi et al. Phys. Rev. Lett. 112, 164802 (2014)
\bibitem{Dattoli24}
G. Dattoli, V. Petrillo, J. V. Rau, 2012 Opt. Commun. 285, 5341
\bibitem{Dattoli25}
G. Dattoli, P.L. Ottaviani and S. Pagnutti, \emph{Booklet for fel design}, Published by ENEA-Edizioni Scientifiche, October 2007.
\bibitem{Maier26}
A. N. Maier et al. Phys. Rev. X 2, 031019 (2012)
\bibitem{Seggenbrock27}
 T. Seggenbrock et al., Phys. Rev. STAB, 16, 070703 (2013)
 \bibitem{Hooker28}
S. Hooker, Rutherford Appleton Laboratory,University of Oxford 9th September 2009
\bibitem{Huang28bis}
Z. Huang, P. Baxevanis, C. Benedetti, Y. Ding, R. Ruth, C. Schroeder, D.Wang and T. Zhang,\emph{Compact laser-plasma-accelerator-driven
free-electron laser using a transverse gradient undulator},Contributed to at the 16th Advanced Accelerator Concepts Workshop San Jose, CA,
USA July 13 - 18, 2014
\bibitem{Ciocci28tris}
F. Ciocci, G. Dattoli and E. Sabia, \emph{Transverse Gradient Undulators and FEL operating with large energy spread: A review}, submitted for publication
\bibitem{Huang29}
Z. Huang et al. Phys. Rev. Lett. 109, 204801 (2012) \\
T. I. Smith, L.R. Elias, J. M. J. Madey, D. A. G. Deacon, J. Appl. Phys. 50, 4580 (1979) \\
The FEL dynamics with TGU has been recently considered by F. Ciocci, G. Dattoli, E. Sabia, to be published. It has been shown that particular care should be devoted to the design of the undulator itself, because  the transverse focusing terms, if not properly adjusted, may introduce inhomogeneous broadenings vanishing the beneficial effects  associated with the transverse gradient itself.
\bibitem{Pukhov30}
A. Pukhov and S. Gordienko, Phil. Transactions Royal Soc. A 364,  623 (2006)
\bibitem{Fowler31}
J. E. Fowler et al. J. Phys. D: Appl. Phys. 46, 325501 (2013)
\bibitem{Shumail32}
M. Shumail et al., in Proc. IPAC 2011 (European Physical Society Accelerator Group, San Sebastian Spain 2011) p.3326.
\bibitem{Dattoli33}
G. Dattoli and P. L. Ottaviani,  Design considerations for X-ray free electron lasers J. Appl. Phys.86, 5331-5336 (1999).
\bibitem{stupakov34}
G. Stupakov, \emph{Using the Beam-Echo Effect for Generation of Short-Wavelength Radiation},Phys. Rev. Lett. Phys. Rev. Lett. 102, 074801 - Published 17 February 2009.
\bibitem{Dattoli35}
G. Dattoli, L. Gianessi, P.L. Ottaviani, H.P. Freund, S.G.Biedron, and S.V. Milton, \emph{Two Harmonic Undulators and Harmonic Generation in High Gain Free Electron Lasers}, Nucl. Instr. Meth. Phys. Res. A 495, 48 (2002).




\end{thebibliography}
\end{document}